\documentclass[prb,amsfonts,amssymb,floats,twocolumn,aps,floatfix,longbibliography,10pt, nofootinbib,groupedaddress,superscriptaddress]{revtex4-2}
\usepackage{graphicx}
\usepackage{graphics}
\usepackage{amsmath}
\usepackage{amssymb}
\usepackage{amsfonts}
\usepackage{dsfont}
\usepackage{braket}
\usepackage{color}
\usepackage{braket,slashed}
\usepackage[mathscr]{euscript}
\definecolor{darkblue}{rgb}{0, 0, 0.8}
\usepackage[colorlinks=true, breaklinks=true, linkcolor=red, citecolor=blue, urlcolor=blue]{hyperref} 
\usepackage{hyperref}
\usepackage{subfigure}
\usepackage{xfrac}
\usepackage{bm}
\usepackage{kantlipsum}
\usepackage{enumitem}
\usepackage{tikz}
\usepackage{framed}
\usepackage{graphicx}
\usepackage{subfigure}
\usepackage{cleveref}
\usepackage{array}

\allowdisplaybreaks[1]

\newcommand{\code}[1]{\texttt{#1}}

\newcommand{\mT}{\ensuremath{\mathcal{T}}}
\newcommand{\mR}{\ensuremath{\mathcal{R}}}
\newcommand{\mQ}{\ensuremath{\mathcal{Q}}}

\newcommand{\hc}{\ensuremath{\mathrm{h.c.}}}

\newcommand{\one}{\mathds{1}}

\renewcommand{\dag}{^\dagger}
\newcommand{\pdag}{^{\phantom{\dagger}}}

\renewcommand{\d}{\ensuremath{\mathrm{d}}}
\newcommand{\e}{\ensuremath{\mathrm{e}}}

\renewcommand{\vr}{\ensuremath{\vec{r}}}

\newcommand{\three}{\ensuremath{\mathbf{3}}}

\newcommand{\eight}{\ensuremath{\mathbf{8}}}

\newcommand{\SU}{\ensuremath{\mathrm{SU}}}
\newcommand{\Z}{\ensuremath{\mathbb{Z}}}
\newcommand{\U}{\ensuremath{\mathrm{U}}}
\newcommand{\OO}{\ensuremath{\mathrm{O}}}

\newcommand{\mC}{\ensuremath{{\mathcal{C}}}}

\renewcommand{\one}[0]{\ensuremath{(\mathbf{1}})}
\newcommand{\two}[0]{\ensuremath{(\mathbf{2}})}
\renewcommand{\three}[0]{\ensuremath{(\mathbf{3}})}
\newcommand{\four}[0]{\ensuremath{(\mathbf{4}})}
\newcommand{\five}[0]{\ensuremath{(\mathbf{5}})}
\newcommand{\six}[0]{\ensuremath{(\mathbf{6}})}
\newcommand{\seven}[0]{\ensuremath{(\mathbf{7}})}
\renewcommand{\eight}[0]{\ensuremath{(\mathbf{8}})}
\newcommand{\nine}[0]{\ensuremath{(\mathbf{9}})}
\newcommand{\oplusL}[0]{\ensuremath{\hspace{-0.1em}\oplus\hspace{-0.1em}}}

\begin{document}

\title{Fractional Chern insulators on cylinders: Tao-Thouless states and beyond}

\author{Felix A. Palm}
\author{Chloé Van Bastelaere}
\affiliation{Center for Nonlinear Phenomena and Complex Systems, Universit\'e Libre de Bruxelles, CP 231, Campus Plaine, 1050 Brussels, Belgium}
\affiliation{International Solvay Institutes, 1050 Brussels, Belgium}

\author{Laurens Vanderstraeten}
\email{laurens.vanderstraeten@ulb.be}
\affiliation{Center for Nonlinear Phenomena and Complex Systems, Universit\'e Libre de Bruxelles, CP 231, Campus Plaine, 1050 Brussels, Belgium}

\date{\today}

\begin{abstract}
Topological phases in two-dimensional quantum lattice models are often studied on cylinders for revealing different topological properties and making the problem numerically tractable. This makes a proper understanding of finite-circumference effects crucial for reliably extrapolating the results to the thermodynamic limit. Using matrix product states, we investigate these effects for the Laughlin-1/2 phase in the Hofstadter-Bose-Hubbard model, which can be viewed as the lattice discretization of the bosonic quantum Hall problem in the continuum. We propose a scaling of the model's parameters with the cylinder circumference that simultaneously approaches the continuum and thermodynamic limits. We find that different scaling schemes yield distinct topological signatures: we either retrieve a spontaneous formation of charge density wave ordering reminiscent of the Tao-Thouless states, known from the continuum problem on thin cylinders, or we find uniform states with a topological degeneracy that can be identified as minimally entangled states known from studies of chiral spin liquids on cylinders. Finally, we carry out a similar analysis of the non-Abelian Moore-Read phase in the same model. Our results clarify the role of symmetries in numerical studies of topologically ordered states on cylinders and highlight the role of lattice effects.
\end{abstract}

\maketitle

\section{Introduction}

The Hofstadter-Bose-Hubbard model~\cite{Harper1955, Azbel1964, Hofstadter1976} describes bosons hopping on a two-dimensional (2D) lattice under a uniform magnetic field with on-site interactions. Due to the interplay of the topological band structure and the interactions, the model is expected to host many exotic quantum phases. In the continuum limit, the model maps to a bosonic quantum Hall system, known to host bosonic versions of fractional quantum Hall (FQH) states~\cite{Wilkin1998, Cooper1999, Wilkin2000, Cooper2001, Paredes2001, Rezayi2005}. In the last twenty years, the effect of the lattice has been investigated in many different contexts, as the lattice can both lead to richer physics and drive the system out of a topological phase due to dispersion broadening the flat band.

Because of the strong effect of interactions, one has to take recourse to numerical methods for mapping out the phase diagram. Numerical explorations of the model, or closely related fermionic versions, range from exact diagonalizations on small clusters~\cite{Sorensen2005, Hafezi2007, Laeuchli2013, Moller2015, Andrews2018}, tensor network approaches such as matrix product states on cylinders and strips~\cite{Zhu2015, Motruk2015, Motruk2015b, He2017, Motruk2017, Dong2018, Rosson2019, Schoonderwoerd2019, Palm2021, Andrews2021, Boesl2022, Palm2025, Yang2025, VanBastelaere2025}, tree tensor networks~\cite{Gerster2017, Macaluso2020}, and projected entangled-pair states on the infinite plane~\cite{Weerda2024}, or neural quantum states~\cite{Doeschl2025, Ledinauskas2025}. These studies have revealed the presence of different types of fractional Chern insulators (FCI), the lattice equivalent of FQH states~\cite{Moller2009, Regnault2011, Sun2011, Neupert2011, Tang2011, Sheng2011, Scaffidi2012, Bergholtz2013, Parameswaran2013}. These topological phases are identified by their signatures in the entanglement entropy~\cite{Kitaev2006, Levin2006} and the entanglement spectrum~\cite{Li2008, Katsura2012}, the fractionalization of quasiparticle excitations~\cite{Laughlin1983, Halperin1984, Arovas1984}, the fidelity with trial topological wavefunctions~\cite{Laughlin1983, Moore1991}, a non-trivial many-body Chern number~\cite{Niu1985, Tao1986}, or the display of chiral edge modes~\cite{Wen1990}.

\par The elongated cylinder geometry is particularly appealing to map out the model's bulk properties in a numerical simulation. In the context of the fermionic FQH effect in the continuum, it was noted numerically that charge density wave (CDW) ordering starts to appear on narrow cylinders~\cite{Rezayi1994}. This connects to an early proposal of Tao and Thouless~\cite{Tao1983} for a wavefunction with crystalline order to describe the ground state of the $\nu=1/3$ FQH system. Despite some initial attempts to explain certain features of the FQH effect~\cite{Anderson1983, Tao1984, Su1984, Su1985}, the proposal was soon abandoned~\cite{Thouless1985} in light of the success of the Laughlin wavefunction~\cite{Laughlin1983}. Two decades later, however, the Tao-Thouless procedure was revisited~\cite{Seidel2005, Bergholtz2005, Bergholtz2006, Seidel2006, Bergholtz2008}: An effective one-dimensional (1D) model can be derived in the thin cylinder limit for which the crystalline Tao-Thouless states are the exact ground states. Moreover, it was shown numerically that this limit is adiabatically connected to the 2D FQH states, so that the Tao-Thouless states can be seen as the quasi-1D precursors of topologically ordered states in 2D. 

On the lattice, the elongated cylinder geometry has been used extensively in numerical simulations, where the formation of CDW order has also been studied and observed~\cite{Bernevig2012, Grusdt2014, Palm2021, Michen2023}. Nonetheless, a systematic understanding is lacking of when and why the spontaneous breaking of translation symmetry occurs on the lattice, and how it scales with cylinder circumference. Interestingly, the effective 1D mapping that leads to the Tao-Thouless states can be used to adiabatically connect FCI states on the lattice to their continuum equivalents~\cite{Qi2011, Scaffidi2012, Wu2012}. This suggests that the analysis in terms of Tao-Thouless states can also be applied to lattices.

\par In this work, we will systematically investigate both lattice and finite-circumference effects in the Hofstadter-Bose-Hubbard model on infinite cylinders by means of matrix product state (MPS) simulations. To this end, we will propose a scaling of the model's parameters with the cylinder circumference, so that we approach both the continuum limit and the thermodynamic limit at the same time. Focusing on the bosonic Laughlin-1/2 phase, we will consider two possible choices for this scaling approach; the first choice directly leads to the lattice equivalent of the Tao-Thouless states, whereas the second choice leads to an analysis in terms of minimally entangled states. Here we will compare the entanglement spectrum of the MPS ground-state approximation with the characteristic level counting in terms of conformal field theories (CFT), as the main signature of topological order. Finally, we will also investigate the topological signatures of the Moore-Read state in the same model.

\section{Tao-Thouless states in the continuum FQH effect}
\label{sec:tt}

\begin{figure}
    \centering
    \includegraphics[width=\columnwidth]{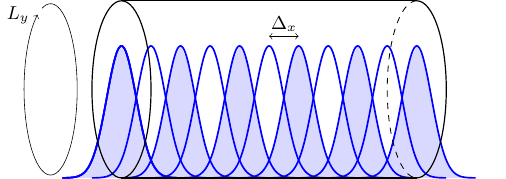}
    \caption{Sketch of the Tao-Thouless state with filling $\nu=1/2$ on the cylinder with circumference $L_y$. The one-particle orbitals $\varphi_n$ are separated by a distance $\Delta_x$; in the Tao-Thouless state, the shaded orbitals are filled and the white ones are empty.}
    \label{fig:tt}
\end{figure}

Let us first revisit interacting 2D bosons in the continuum under a uniform perpendicular magnetic field $B$, described by the Hamiltonian
\begin{multline} \label{eq:ham_cont}
    H = \int d\vec{r} \; \bigg( \psi\dag(\vec{r}) \left(  i \vec{\nabla} -  \vec{A}(\vr) \right)^2 \psi(\vec{r}) \\ + g_2 \;  \rho(\vr) \rho(\vr)  \bigg),
\end{multline}
where $\vec{A}(\vec{r})$ is a vector potential for the magnetic field $B$. Here we have introduced the bosonic field operator $\psi(\vr)$ and $\rho(\vr)=\psi\dag(\vec{r})\psi(\vec{r})$ is the density operator. We express the magnetic field in units of the elementary magnetic flux $\Phi_0$, so that the magnetic length $\ell$ is given by 
\begin{equation} \label{eq:ell}
    \ell = \sqrt{\frac{1}{2\pi B}}.
\end{equation}
We consider a cylindrical geometry with coordinates $(x,y)$, where $x$ runs along the elongated direction and $y$ winds around the cylinder (see Fig.~\ref{fig:tt}). We take the limit of an infinite cylinder, $L_x\to\infty$, with a finite circumference $L_y$. Choosing a Landau gauge
\begin{equation}
    \vec{A}(x,y) = \left( 0, Bx \right),
\end{equation}
the system is translation invariant in the $y$ direction, and the translation symmetry in the $x$ direction is broken into discrete translations $x\to x + n \Delta_x$ with a period
\begin{equation}
    \Delta_x = \frac{2\pi \ell^2}{L_y} \;.
\end{equation}
The one-particle orbitals $\varphi_n$ in the lowest Landau level,
\begin{equation} \label{eq:orbitals}
    \varphi_n(x,y) \propto \e^{ik_n y} \exp\left(- \frac{1}{2\ell^2} \left( x-n \Delta_x \right)^2 \right),
\end{equation}
have well-defined momentum along $y$, 
\begin{equation}
    k_n = \frac{2\pi}{L_y}n
\end{equation}
and are now centered around equally spaced locations $n\Delta_x$ along the cylinder (see Fig.~\ref{fig:tt}). When we project an interaction term into this space of lowest Landau level orbitals, we obtain an effective interacting 1D lattice model.

If we now introduce a finite density of bosons, the interaction term stabilizes different types of bosonic FQH states~\cite{Wilkin1998, Cooper1999, Wilkin2000, Cooper2001} at different rational values for the magnetic filling factor $\nu=n_b / B $, with $n_b=\braket{\rho}$ the boson number density. The best known example is the bosonic Laughlin state~\cite{Laughlin1983} at $\nu=1/2$, which appears as the exact ground state of the contact interaction term~\cite{Wilkin1998}. Moreover, it was shown numerically~\cite{Regnault2003, Regnault2004} that there is a finite excitation gap in the thermodynamic limit (the most precise numerical estimate around $\Delta E \approx 0.615(5) \times V_0$~\cite{Repellin2014}), confirming that the Laughlin-1/2 state describes an incompressible phase in the model \eqref{eq:ham_cont}. Here, the gap is expressed in terms of the first pseudo-potential $V_0$~\cite{Haldane1983}, which for contact interactions is the only relevant energy scale in the system and is given by the splitting between the ground state and the first excited state of a system of two interacting bosons in the lowest Landau level.

\par As first pointed out for the fermionic FQH effect~\cite{Rezayi1994}, it is found that for small circumferences the ground state of a FQH system develops density modulations. In the case of the bosonic $\nu=1/2$ phase, the ground state exhibits a charge density wave (CDW) with a wave vector \mbox{$k=\pi/\Delta_x$}, and an associated order parameter 
\begin{equation} \label{eq:cdw_cont}
    m_{\mathrm{CDW}} = \int \d\vec{r} \; \e^{i k x} \braket{\rho(\vec{r})} \;.
\end{equation}
This CDW ordering corresponds to a staggered filling of the Landau orbitals $\varphi_n$. It was, in fact, realized that in the limit $L_y\to0$ the exact ground states are the product states with alternating filled and empty orbitals~\cite{Bergholtz2005, Bergholtz2006, Bergholtz2008} (see Fig.\ref{fig:tt}). These crystalline states, denoted as $[1010]$ and $[0101]$, correspond to the original proposal of Tao and Thouless~\cite{Tao1983}. For finite values of $L_y$, these states get dressed with quantum fluctuations and the CDW order parameter decreases and eventually goes to zero as $L_y\to\infty$. The vanishing of the order parameter was observed~\cite{Seidel2005} to be exponential as
\begin{equation} \label{eq:o_cont}
m_{\mathrm{CDW}} \propto \exp(- L_y^2/\ell^2),\qquad  L_y \gg \ell \;.
\end{equation}

Numerics show that there is an adiabatic path between the Tao-Thouless states on the narrow cylinder and the Laughlin state~\cite{Laughlin1983} in the 2D system. Moreover, these crystalline states exhibit many topological signatures of the fully 2D FQH system, such as ground-state degeneracy, entanglement spectra~\cite{Laeuchli2010, Liu2012} and quasiparticle fractionalization~\cite{Seidel2006, Bergholtz2006b}. In that sense, these Tao-Thouless states can be considered as the quasi-1D precursors of the 2D Laughlin-1/2 state.

\par Note that the spontaneous breaking of translation symmetry for any finite $L_y$ is a direct consequence~\cite{Seidel2005} of Oshikawa's generalization~\cite{Oshikawa1997, Oshikawa2000} of the Lieb-Schultz-Mattis theorem~\cite{Lieb1961}: a gapped lattice system with a conserved $\U(1)$ charge and a $m/n$ filling of the unit cell (with $m$ and $n$ co-prime) must have at least an $n$-fold degenerate ground state. Since we are dealing with an effective 1D model, these $n$ different ground states are distinguishable by a local order parameter, in this case the CDW ordering. In that sense, the formation of crystalline states is an unavoidable consequence of the symmetries in the model, and is also expected to occur in lattice realizations of FQH states on the cylinder.

\section{Lattice model}
\label{sec:model}

Let us now turn to the Hofstadter-Bose-Hubbard model on a square lattice cylinder. In general terms, the Hamiltonian is given by
\begin{equation}
\begin{aligned}
H = & - t \sum_{x=-\infty}^\infty\sum_{y=1}^{N_y} \left( \e^{i\phi_x y} b\dag_{x,y} b\pdag_{x+1,y} + \hc \right) \\
& -t \sum_{x=-\infty}^\infty\sum_{y=1}^{N_y} \left( \e^{i\phi_y x} \e^{i\Phi/N_y} b\dag_{x,y} b\pdag_{x,y+1} + \hc \right) \\ & + \frac{U_2}{2} \sum_{x=-\infty}^\infty\sum_{y=1}^{N_y} n_{x,y} (n_{x,y} -1 ) \\
& + \frac{U_3}{6} \sum_{x=-\infty}^\infty\sum_{y=1}^{N_y} n_{x,y} (n_{x,y} -1 ) (n_{x,y} -2 ) \label{eq:ham} \;,
\end{aligned}
\end{equation}
with $b_{x,y}\dag$ and $b_{x,y}\pdag$ the bosonic creation and annihilation operators at site $(x,y)$, and $n_{x,y}=b_{x,y}\dag b_{x,y}\pdag$ the number operator. The values of the Peierls phases $(\phi_x,\phi_y)$ in the hopping amplitudes depend on the gauge; in this work, we consider both the Landau gauge in the $x$-direction, $(\phi_x,\phi_y)=(2\pi\alpha,0)$, and in the $y$-direction, $(\phi_x,\phi_y)=(0,2\pi\alpha)$, where $\alpha$ is the magnetic flux per plaquette. On the cylinder geometry, the former choice can only be consistent with the periodic boundary conditions if $\alpha$ is a multiple of $1/N_y$. In that case, the gauge choices are related by a local unitary transformation. We introduce a typical on-site Hubbard repulsion term with strength $U_2$. A final parameter of the model is the average boson density $n_b$, which leads to the magnetic filling fraction $\nu=n_b/\alpha$. Finally, for future purposes, we include a three-body repulsion with strength $U_3$ and a magnetic flux $\Phi$ piercing through the cylinder, such that a boson picks up a phase $\Phi$ when hopping around the cylinder once.

We can interpret this model as a lattice discretization of the continuum quantum Hall problem [Eq.\eqref{eq:ham_cont}]. If we introduce the lattice spacing $a$, the above phase factors are related to the magnetic length $\ell$ as
\begin{equation}
    \alpha = \frac{a^2}{2\pi \ell^2},
\end{equation}
where $\ell$ was defined in the continuum according to Eq.~\eqref{eq:ell}. If we want to find a continuum limit ($a\to0$) of our model with a finite value of the magnetic length, the flux per plaquette should approach $\alpha\to0$ while keeping $\alpha/a^2$ constant. We will implement this scaling by considering cylinders with an increasing number of lattice sites $N_y$, while at the same time scaling $\alpha\propto 1/N_y$. The effective size of the cylinder is then given by
\begin{equation} \label{eq:Ly}
    L_y = a N_y \propto \frac{a}{\alpha} = \sqrt{\frac{2\pi\ell^2}{\alpha}},
\end{equation}
meaning we are considering cylinders of increasing size as 
\begin{equation} \label{eq:scaling_Ly}
    \frac{L_y}{\ell} \propto \sqrt{N_y}.
\end{equation}
By scaling $\alpha\propto1/N_y$ with $N_y \to \infty$ in our lattice model, we are therefore considering the continuum limit ($\alpha\to0$) and the thermodynamic limit ($L_y\to\infty$) at the same time. In our simulations we keep the magnetic filling fraction $\nu$ constant, implying that the boson density also scales as $n_b=\nu/N_y$.

\begin{figure}
    \centering
    \includegraphics[width=0.9999\linewidth]{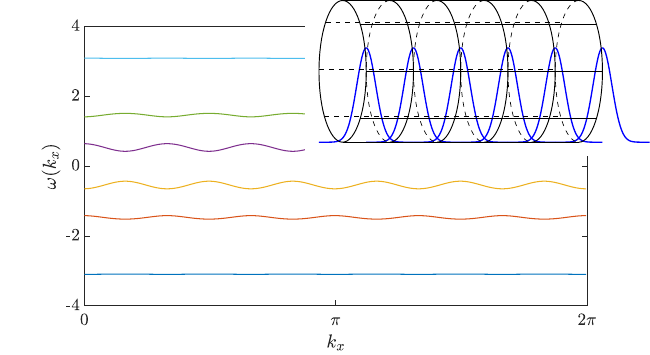}
    \includegraphics[width=0.9999\linewidth]{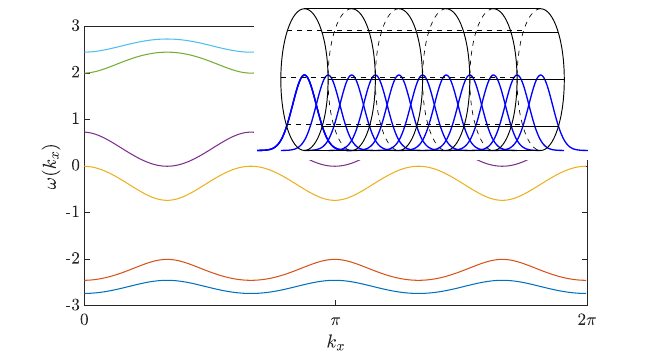}
    \caption{The single-particle band structure of the Hofstadter-Bose-Hubbard model in Eq.~\eqref{eq:ham} with $\alpha=1/N_y$ (top) and $\alpha=2/N_y$ (bottom) for $N_y=6$. The inset shows the single-particle orbitals [Eq.~\eqref{eq:orbitals}] on the cylinder in the continuum, with the associated lattice discretization.}
    \label{fig:bands}
\end{figure}

In the following, we consider two particular choices for carrying out this scaling analysis, $\alpha=1/N_y$ and $\alpha=2/N_y$. These values are particularly convenient as they lead, in the Landau-$x$ gauge, to a fully translation invariant Hamiltonian in the $x$-direction. As compared to the continuum case, the lattice structure gives rise to non-trivial features of the single-particle band structure. In Fig.~\ref{fig:bands} we show the single-particle energies of the lattice Hamiltonian on the cylinder for the two cases $\alpha=1/N_y$ and $\alpha=2/N_y$. Using the translation invariance in the $x$-direction, we plot the dispersion as a function of the corresponding momentum $k_x$. For $\alpha=1/N_y$, we obtain a single low-lying band that is well-separated from the rest of the spectrum, whereas for the case $\alpha=2/N_y$ we find that the two lowest-lying bands get very close together for increasing values of $N_y$\footnote{In fact, for even $N_y$ the lowest two sub-bands touch, because they correspond to the same band but are assigned different values of $k_y$.}. In the inset of the same figure, we visualize how the lattice discretization intersects with the single-particle orbitals $\varphi_n$ [Eq.~\eqref{eq:orbitals}] that were found on the cylinder in the continuum: When we choose $\alpha=1/N_y$, the lattice spacing is exactly equal to the spacing between the different orbitals, whereas for $\alpha=2/N_y$ the lattice spacing is only half of the continuum orbital spacing. 

Connecting back to the discussion of energy scales in the continuum FQH problem, we want to relate the energy scales of the lattice model to the continuum FQH regime. To this end, we identify the characteristic energy scale of the hard-core bosonic two-body problem in analogy to solving the two-body problem in the lowest Landau level. In particular, we calculate the low-lying spectrum for two hard-core bosons at flux \mbox{$\alpha=1/N_y$} on a cylinder of radius $N_y$ and varying length \mbox{$N_x = N_y, \hdots, 15$}. Above the quasi-degenerate ground-state manifold, we find a characteristic gap $\Delta_{\infty}(N_x, N_y)$ to the lowest-lying excited states. The values obtained for the gap from these simulations are in excellent agreement for increasing $N_x$, so that we use their mean value as the characteristic gap $\Delta_{\infty}(N_y)$, which plays the role of the Haldane pseudo-potential $V_0$ in the continuum~\cite{Haldane1983, Laeuchli2013, Repellin2014}. Expressing the energies in the FCI problem in units of $V_0$ therefore allows us to draw a direct connection from the lattice model to its continuum analog. Note that we did not perform any projection to the lowest Hofstadter band, so that band mixing effects from the local repulsion are already accounted for in our simulations.

\section{Methods and symmetries}
\label{sec:mps}

\begin{figure}
    \centering
    \includegraphics[scale=0.4,page=1]{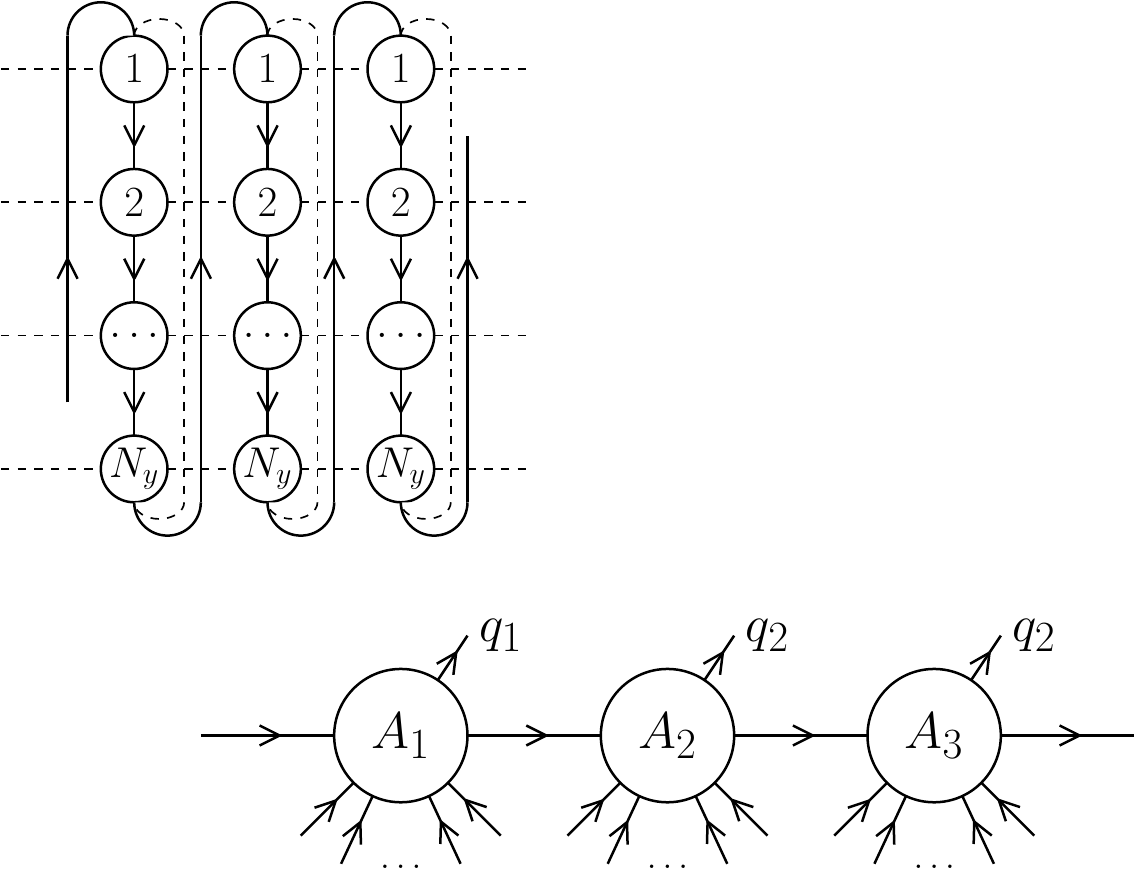}
    \caption{Generic MPS representation of the ground state of the Hofstadter-Bose-Hubbard model [Eq.~\eqref{eq:ham}] on the cylinder. We snake a 1D path through the cylinder, giving rise to an effective 1D model with $N_y$ unit cell. For analyzing the symmetry structure of the MPS, we can group each unit cell or rung in a super-site with $N_y$ physical legs (in the actual numerical optimization we always decompose into different MPS tensors). Depending on the situation, we can consider a multi-rung unit cell, where each MPS tensor $A_i$ carries a corresponding $\U(1)$ charge $q_i$. Note also that the translation operator $\mT_y$ and the reflection operator $\mR_y$ act as the product of operators acting locally as a permutation within the super-site space.}
    \label{fig:mps}
\end{figure}

In this work, we simulate the model in Eq.~\eqref{eq:ham} on infinite cylinders using tensor network methods~\cite{Cirac2021}. To this end, we trace a 1D path around the cylinder, effectively mapping the cylinder geometry to a purely 1D model, see Fig.~\ref{fig:mps}. The ground state can be variationally approximated directly for the infinite cylinder by uniform matrix product states (MPS)~\cite{Vanderstraeten2019}, which we optimize using the VUMPS algorithm~\cite{ZaunerStauber2018}. The MPS bond dimension $D$ serves as the control parameter in our simulations and we have ensured convergence of all results that we report in this work. The bond dimension needed for convergence is expected to scale exponentially with the cylinder size $N_y$ \cite{Stoudenmire2012}, which limits us to $N_y=12$ in this work. For optimizing ground-state MPS, we always work in the Landau-$x$ gauge, where the effective 1D Hamiltonian has a unit cell of size $N_y$. If needed, we can switch to the Landau-$y$ gauge after the optimization by applying a local unitary operation on the MPS.

The model has a global $\U(1)$ symmetry, generated by the number operator
\begin{equation}
    \mQ = \sum_{x,y} n_{x,y} \; .
\end{equation}
Since we are only interested in fillings that are commensurate with $1/N_y$, we can straightforwardly fix the particle density $n_b$ by explicitly encoding the $\U(1)$ symmetry of the model into the MPS representation. The resulting structure of the MPS is represented graphically in Fig.~\ref{fig:mps}. In this context, the Lieb-Schultz-Mattis-Oshikawa theorem~\cite{Lieb1961, Oshikawa1997, Oshikawa2000} can be found as a direct consequence of the structure of the $\U(1)$ charges in an MPS ground state~\cite{Sanz2009, Cirac2021}. If the average boson density is not commensurate with the unit cell we are considering, we have to break translation symmetry in the MPS explicitly by choosing different $\U(1)$ charges $q_i$ on consecutive tensors $A_i$ within a larger unit cell. For a density $m/n$ per rung (with $m$ and $n$ co-prime), we have to impose an $n$-site unit cell. Alternatively, we can impose a uniform fractional charge $q_i=m/n$, but this will lead to a staggering of the fractional charges on the virtual MPS bonds, also leading to an $n$-site unit cell. We will see an explicit example for a 1/2 charge per rung in Sec.~\ref{sec:ln1}, see Fig.~\ref{fig:mps_ln1}.

Apart from the global $\U(1)$ symmetry, there are also a number of spatial symmetries, which depend on the gauge for the magnetic field. With the Landau-$x$ gauge, the model has translation symmetry $\mathcal{T}_x$ along $x$, whereas in the Landau-$y$ gauge, $\mathcal{T}_y$ is a symmetry of the model. Without an external flux ($\Phi=0$), the model is also symmetric under the combined action of time reversal and reflection in the $x$-direction and the $y$-direction; because of the anti-unitary nature of the time-reversal symmetry, we will refer to these as anti-unitary reflection symmetries $\mR_{x/y}$. We will consider bond-centered and site-centered reflection symmetries.

Besides local expectation values, the MPS representation gives us direct access to the entanglement spectrum. Here, we focus on bipartite entanglement cuts along the $y$-direction of the cylinder. For characterizing the spectrum, we work in the Landau-$y$ gauge that explicitly preserves $\mT_y$ so that we can label the entanglement eigenvalues with charge and $y$-momentum quantum numbers. The Schmidt decomposition then reads 
\begin{equation} \label{eq:es}
    \ket{\Psi} = \sum_{q,k_y} \sum_i \lambda^i_{q,k_y} \ket{\Psi^{L,i}_{+q,k_y}} \otimes \ket{\Psi^{R,i}_{-q,k_y}},
\end{equation}
with positive Schmidt values $\lambda^i_{q,k_y}$. The labels for the left/right Schmidt states denote that these are eigenstates of the charge and translation operator acting to the left/the right of the entanglement cut with eigenvalues $(q,k_y)$. We can extract the $\U(1)$ quantum number of each Schmidt value directly from the $\U(1)$ symmetry of the MPS representation, but the charges $q$ are only defined up to a global reference value: All the charge labels can be shifted by adding a charge at one of the boundaries, without observable consequences in the thermodynamic limit. This means we can always shift the charge labels in order to make the physical interpretation more transparent. Similarly, the value of $k_y$ is only defined up to a global shift, which we also fix by shifting the dominant Schmidt values to $k_y=0$. Note that the snake-like structure of the MPS representation breaks the $\mT_y$ symmetry explicitly, and we have to find the $k_y$ labels of the Schmidt values approximately by tracking the action of the $\mT_y$ operator on the MPS~\cite{Cincio2013}. In practice, for large enough bond dimensions, $\mT_y$ is preserved to very high precision, and we can associate $(q,k_y)$ quantum numbers to the low-lying part of the entanglement spectrum quasi perfectly.

Based on the charge-resolved entanglement spectrum defined above, we can also directly compute the charge difference between the half-infinite regions to the left and the right of the entanglement cut. Indeed, upon defining $Q_{L/R}$ as the charge counting operator acting on the left and the right of a given cut, we find
\begin{equation} \label{eq:deltaQ}
    \Delta Q = \bra{\Psi}\frac{Q_L-Q_R}{2}\ket{\Psi} = \sum_q \sum_i(\lambda^i_q)^2 q .
\end{equation}
Note, again, that in the thermodynamic limit the charges of the Schmidt values are only defined up to a global shift, so only relative values for $\Delta Q$ are unambiguously defined.

The bond-centered reflection symmetry considered above leaves an important fingerprint in the charge-resolved entanglement spectrum. Acting with this symmetry on the Schmidt decomposition in Eq.~\eqref{eq:es} maps every term with charge $q$ to a term with charge $-q$, implying that all Schmidt values $\lambda^i_q$ for $q\neq0$ are necessarily degenerate\footnote{Possibly we need to first shift all the charge labels to make this symmetry apparent.}. Here we can differentiate two situations: either the entanglement spectrum has values with $q=0$ (which are not degenerate) and all other values $q=\pm1,\pm2,\dots$ are symmetric under $q\to-q$, or the entanglement spectrum consists entirely of twofold degenerate Schmidt values that can be made symmetric by shifting the labels to half-integer labels $q=\pm\frac{1}{2},\pm\frac{3}{2}, \dots$. We interpret this as signs of an effective $\OO(2)$ symmetry of the entanglement spectrum, originating from the semi-direct product of the $\U(1)$ symmetry and the charge-conjugation symmetry. In the former case of integer-valued charges the entanglement spectrum follows the integer representations of $\OO(2)$ with labels $n=0,1,2,\dots$ (for $n=0$ there are two one-dimensional irreducible representations, all higher $n$ denote two-dimensional irreducible representations), whereas in the latter case the spectrum follows the projective representations of $\OO(2)$, labeled as $n=\frac{1}{2},\frac{3}{2},\dots$ (which are all two-dimensional).

\section{Laughlin-1/2 phase, first choice}
\label{sec:ln1}

\begin{figure}
    \centering
    \includegraphics[width=0.9999\columnwidth
    ,page=2]{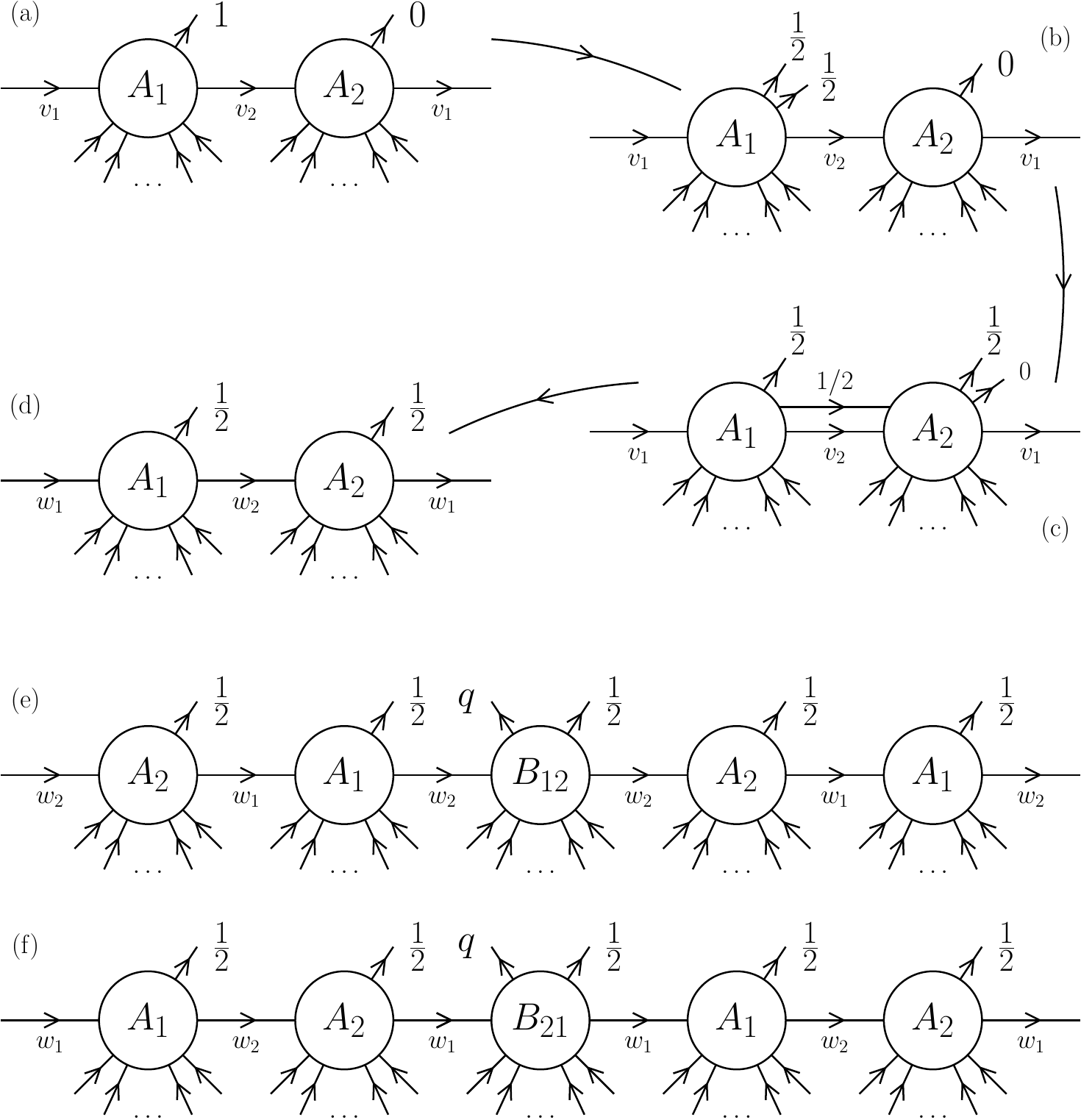}
    \caption{Top: the equivalence between two different ways of structuring the U(1) charges in the MPS, where we have used the super-site representation as in Fig.~\ref{fig:mps}. We start from the MPS representation where we put integer charge on one of the tensors (a), we split this charge into half (b), we drag this half-integer charge unto the next tensor (c), and fuse the double-line leg by shifting all the virtual charges by an amount of $1/2$ (d). Bottom: the two quasiparticle configurations. We start from the MPS representation in (d), and make a defect between tensors $A_1$ and $A_2$ (e) or vice-versa (f). The corresponding tensors $B_{12}$ and $B_{21}$ carry an extra charge label $q$, which denotes the global charge of the excitation relative to the ground state charge and has to be half-integer. In the excitation ansatz, we make another momentum superposition of such a configuration, and variationally optimize over the tensors $B_{12}$ or $B_{21}$.}
    \label{fig:mps_ln1}
\end{figure}

We first focus on searching for topological signatures of the bosonic Laughlin state~\cite{Laughlin1983} at a filling factor $\nu=1/2$. We work with hard-core bosons ($U_2\to\infty$), which is implemented by imposing the maximum number of bosons per site to one. As explained in Sec.~\ref{sec:model}, we simulate different circumferences $N_y$ and scale the flux per plaquette accordingly. In this section, we consider the choice
\begin{equation}
    \alpha=\frac{1}{N_y}, \qquad n_b=\frac{1}{2N_y}
\end{equation}
giving rise to $\nu=1/2$. Since we have an effective 1D system with a filling of 1/2 per unit cell, the Lieb-Schultz-Mattis-Oshikawa theorem dictates that we cannot have a unique gapped ground state. This is also reflected in the MPS-ansatz for describing the ground state, using a two-rung unit cell to accommodate a density of 1/2 per rung. We can either place a unit charge on one of the two rungs in the unit cell, or place half a charge on every rung in the unit cell. In the latter case, we have to introduce fractional $\U(1)$ charges, and the virtual MPS charges alternate between integer and half-integer charges. In Fig.~\ref{fig:mps_ln1} it is shown that one can transform the one into the other, and that both choices are entirely equivalent.

\par Upon optimizing this two-rung unit cell MPS, we indeed find a CDW ordering with a period of two sites in the $x$-direction. We define the CDW order parameter as the lattice equivalent of Eq.~\eqref{eq:cdw_cont},  
\begin{equation} \label{eq:cdw}
    m_{\mathrm{CDW}} = \left| \sum_{y=1}^{N_y} \braket{n_{x,y} - n_{x+1,y}} \right|,
\end{equation}
and plot it as a function of $N_y$ in the top panel of Fig.~\ref{fig:ln1}. We find that the magnitude $m_{\rm CDW}$ drops exponentially as the circumference $N_y$ of the cylinder increases. This agrees with the continuum case [Eq.~\eqref{eq:o_cont}] and the effective scaling of the cylinder circumference [Eq.~\eqref{eq:Ly}].

\subsection{Entanglement structure}

\begin{figure}
    \centering
    \includegraphics[width=0.999\columnwidth]{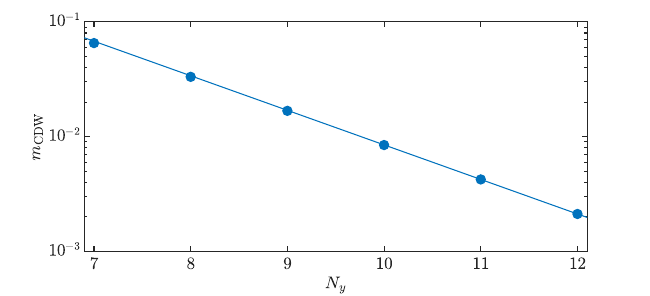}
    \includegraphics[width=0.999\columnwidth]{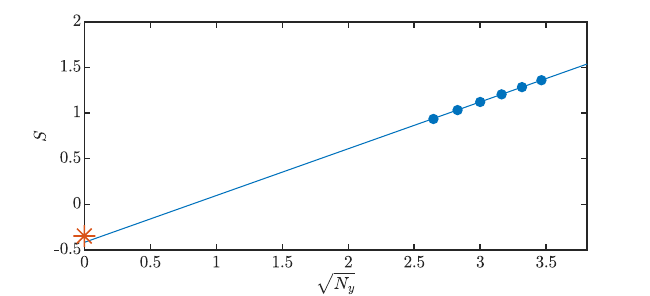}
    \caption{Scaling of the charge density wave order parameter (top) and the entanglement entropy (bottom) for $(\alpha,n_b)=(1/N_y,1/(2N_y))$ as a function of $N_y$. For the former, we fit to the expected exponential decay as a function of $N_y$. For the latter, we fit to a linear behavior as a function of $\sqrt{N_y}$ giving us a value for the topological entanglement entropy $\gamma\approx0.415$.}
    \label{fig:ln1}
\end{figure}

Next, we look at the entanglement entropy corresponding to a bipartition of the cylinder. We find that the entanglement entropies are identical for both entanglement cuts within the unit cell, which results from the charge conjugation symmetry in the entanglement spectra (see below). Since we are varying the parameter $\alpha$ as a function of the cylinder circumference $N_y$, we cannot perform the usual $N_y$-scaling of the entanglement entropy. Instead, we will use that by varying $N_y$ we also approach the continuum limit, and that the effective scaling of the continuum cylinder circumference $L_y$ goes as $L_y/a \propto \sqrt{N_y}$ (see Eq.~\eqref{eq:Ly}). If we plot the value of the entanglement entropy as a function of $\sqrt{N_y}$, we expect to retrieve the scaling
\begin{equation} \label{eq:scaling_S}
    S = c \; \sqrt{N_y} - \gamma + \dots,
\end{equation}
where $c$ is a non-universal proportionality constant and $\gamma$ is the universal value known as the topological entanglement entropy. We confirm this scaling in the bottom panel of Fig.~\ref{fig:ln1}, with the extracted topological entanglement entropy close to its expected value $\gamma=\log\sqrt{2}$. The agreement with the expected scaling for the entanglement entropy [Eq.~\eqref{eq:scaling_S}] validates our procedure of taking the continuum and thermodynamic limits simultaneously. Note that this scaling with the square root of the cylinder circumference is equivalent to expressing the latter in units of the magnetic length, as done in previous works~\cite{Schoonderwoerd2019, Boesl2022}.

\begin{figure}
    \centering
    \includegraphics[width=0.999\columnwidth]{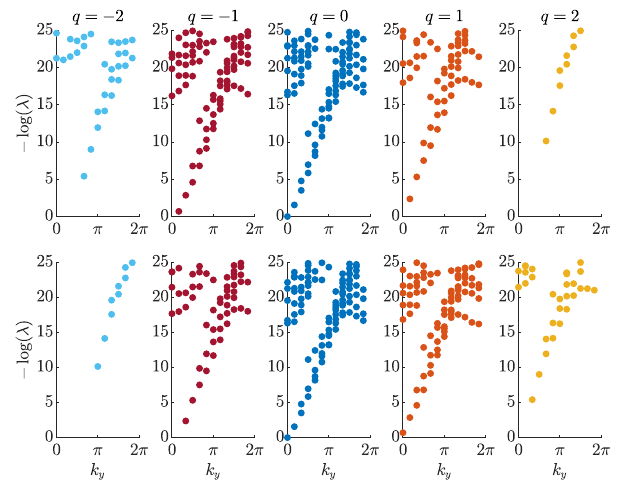}
    \caption{The entanglement spectra for the two different cuts in the two-rung MPS approximation for the case $(\alpha,n_b)=(1/N_y,1/(2N_y))$ with $N_y=12$. Here the $q$ labels refer to the $\U(1)$ charges corresponding to the staggered charge distribution with only integer charges on the virtual MPS legs, see Fig.~\ref{fig:mps_ln1}(a).}
    \label{fig:es_ln1_split}
\end{figure}

Going beyond the entanglement entropy, we now consider the full symmetry-resolved entanglement spectrum. We use the labeling $(q, k_y)$ corresponding to a staggered charge distribution [1010] in the MPS representation (Fig.~\ref{fig:mps_ln1}(a)), such that only integer charges appear on the virtual MPS level. The two different spectra are shown in Fig.~\ref{fig:es_ln1_split}. In each charge sector, we can see a chiral branch emerging with a clear counting $(1,1,2,3,5,7,11,\dots)$, in agreement with the $\SU(2)_1$ CFT counting that is expected for a $\nu=1/2$ Laughlin phase (see App.~\ref{sec:app} for more details). A striking feature of these two entanglement spectra is that they are related by a charge conjugation symmetry $q\to-q$, up to a shift in the $k_y$ labels. This symmetry reflects the fact that the pattern of charge fluctuations in the MPS ground state is symmetric with respect to site-centered reflections in the $x$ direction, which maps a charge-resolved Schmidt decomposition on an even link to the same decomposition on an odd link with conjugated charges (see Sec.~\ref{sec:mps}).

\begin{figure}
    \centering
    \includegraphics[width=0.999\columnwidth]{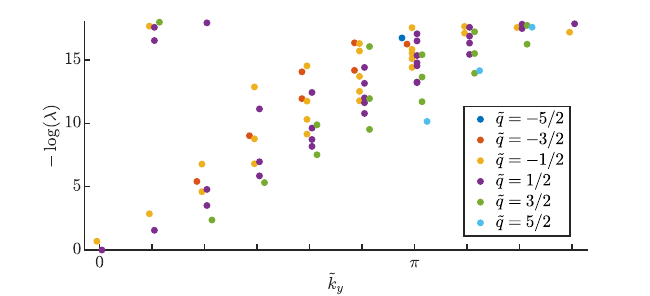}
    \includegraphics[width=0.999\columnwidth]{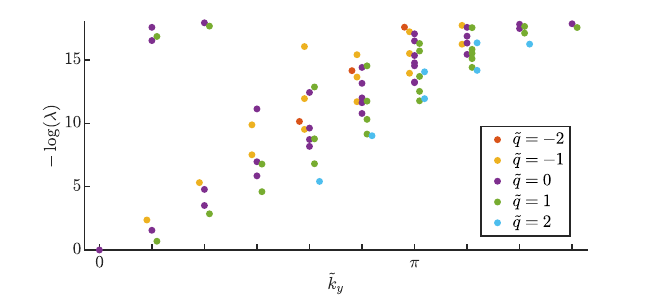}
    \caption{The entanglement spectra for the two different cuts in the two-rung MPS representation for the case $(\alpha,n_b)=(1/N_y,1/(2N_y))$ with $N_y=12$. Here the $q$ labels now refer to the $\U(1)$ charges corresponding to the uniform charge distribution with both integer and half-integer charges, see Fig.~\ref{fig:mps_ln1}(d), whereas the spectrum is plotted as a function of the shifted momentum $\tilde{k}_y$, see Eq.~\eqref{eq:ky_tilde}.}
    \label{fig:es_ln1_comb}
\end{figure}

Finally, we now combine the entanglement spectra in the different $q$ sectors into a single plot. To resolve the quantum numbers of the Schmidt values, we first switch to the MPS representation with uniform charge distribution (Fig.~\ref{fig:mps_ln1}(d)), in which the charge labels of the entanglement spectra interchange between half-integer and integer representations. We label the shifted charges as $\tilde{q}$. Second, we plot the spectrum as a function of a shifted $y$-momentum that is obtained as
\begin{equation} \label{eq:ky_tilde}
    \tilde{k}_y = k_y - q \frac{2\pi}{N_y} \; .
\end{equation}
This label $\tilde{k}_y$ is inspired by the definition of the scaled momentum operator in the FQH setting on the cylinder~\cite{Zaletel2013, Zaletel2015}. Plotting the spectra in the different charge sectors for the two entanglement cuts, we obtain the two panels in Fig.~\ref{fig:es_ln1_comb}. We observe that the level counting now follows the full $\SU(2)$ multiplet structure of the half-integer (i.e., $2,2,6,\dots$) and the integer (i.e., $1,3,4,\dots$) sectors of the $\SU(2)_1$ CFT counting (see App.~\ref{sec:app}). However, we also observe large splittings within the multiplets, which are an unavoidable feature in this set-up: The entanglement eigenvalues have exactly the same magnitude in both sectors, but are just organized differently in their labels $(\tilde{q},\tilde{k}_y)$.

\subsection{Quantized charge pumping}
\label{sec:charge_pumping_ln1}

To further investigate the topological nature of the degenerate ground states, we perform a charge-pumping experiment to extract the many-body Chern number~\cite{Laughlin1983, Zaletel2014, Grushin2015}. Starting from one of the two degenerate ground states, we adiabatically insert a flux quantum through the cylinder by changing $\Phi$ from $0$ to $2\pi$ [Eq.~\eqref{eq:ham}]. We monitor the accumulation of charge on the left or the right of a given bipartition of the system by extracting the left-right charge imbalance from the entanglement spectrum according to Eq.~\eqref{eq:deltaQ}. In particular, by monitoring the entanglement spectrum as a function of $\Phi$ we can obtain the transported charge
\begin{equation}
    Q_{\rm tr} = \Delta Q_{\Phi=2\pi} - \Delta Q_{\Phi=0} + q_{\rm shift},
    \label{eq:transportedCharge}
\end{equation}
where $q_{\rm shift}$ is a shift in the $q$-label of the Schmidt values at the end of the flux insertion~\cite{Zaletel2014} and $\Delta Q$ was defined for a given entanglement spectrum in Eq.~\eqref{eq:deltaQ}.

\begin{figure}
    \centering
    \includegraphics[width=0.999\columnwidth]{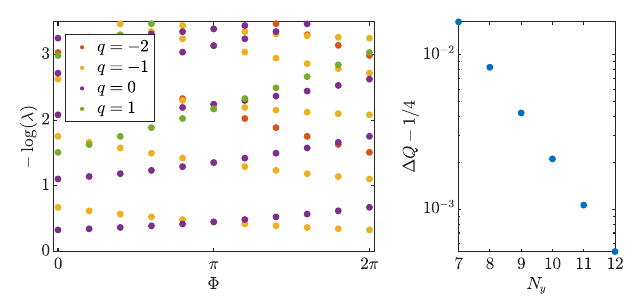}
    \caption{Left: Flux insertion protocol for the case $(\alpha,n_b)=(1/N_y,1/(2N_y))$ at $N_y=12$, showing the charge-resolved entanglement spectrum along the adiabatic tuning of $\Phi$. Right: The left-right charge imbalance of the entanglement spectrum at $\Phi=0$ in the MPS approximation, computed from the entanglement spectra in Fig.~\ref{fig:es_ln1_split} with Eq.~\eqref{eq:deltaQ}, as a function of $N_y$.}
    \label{fig:flux_ln1}
\end{figure}

In Fig.~\ref{fig:flux_ln1}, we show the charge-resolved entanglement spectrum of the MPS ground state as a function of $\Phi$. We start from a spectrum with the leading Schmidt value in the $q=0$ sector and with the first sub-leading value in the $q=-1$ sector, i.e., the spectrum plotted in the top panel of Fig.~\ref{fig:es_ln1_split}. Following the state adiabatically, we evolve into a spectrum with the two leading Schmidt values in the $q=-1$ and $q=0$ sectors, i.e., the spectrum in the bottom panel of Fig.~\ref{fig:es_ln1_split}, but with all the charge labels shifted by $q_{\mathrm{shift}}=-1$. Since the two spectra are related through charge conjugation, we know that $\Delta Q_{\Phi=2\pi}= - \Delta Q_{\Phi=0} \equiv -\Delta Q$, resulting in the transported charge $Q_{\rm tr} = -2\Delta Q + 1$, with $\Delta Q$ obtained from either of the entanglement spectra. In the right panel of Fig.~\ref{fig:flux_ln1} we show that $\Delta Q$ converges to the value $1/4$ exponentially with $N_y$, which corresponds to a transported charge $Q_{\rm tr}=1/2$, and hence a many-body Chern number $\mathcal{C}=1/2$, confirming the topological nature of the state.

\subsection{Low-energy excitations}

Finally, we study the excitation spectrum on top of the MPS ground states. As was observed on the level of the Tao-Thouless states for the continuum case~\cite{Seidel2005, Bergholtz2005}, the elementary excitations on top of the CDW-ordered ground states are domain walls between the two CDW patterns. Using the MPS representations of the two ground states, we can similarly consider domain-wall excitations: Starting from the MPS representation with the uniform charge distribution with interchanging half-integer and integer virtual $\U(1)$ charges, it follows that the charge of the domain wall states has to be half-integer (see Fig.~\ref{fig:mps_ln1}). These fractional charges can be directly related to the quasiparticle and quasihole states on top of the Laughlin-1/2 state. We can variationally target these fractionalized domain wall states with well-defined momentum by applying the MPS excitation ansatz~\cite{Haegeman2012, Vanderstraeten2019, ZaunerStauber2018b, VanDamme2021, Vanderstraeten2020}, yielding a momentum-resolved dispersion relation for the quasiparticle (with charge $q=1/2$) and quasihole (with charge $q=-1/2$).\footnote{As shown in Fig.~\ref{fig:mps_ln1}, the MPS representation of the domain walls admits two choices for the domain walls (denoted by $B_{12}$ and $B_{21}$). For technical reasons the excitation energy of each of these states is only defined up to a global shift, which acts with a different sign in the two cases. Therefore, we can define a quasiparticle and quasihole dispersion by taking the average of both excitation energies.} Besides these domain wall states, we can also consider neutral excitations that are only a local perturbation of one of the two ground states. Here, we can again consider a momentum superposition and variationally optimize in order to find a variational dispersion relation in the neutral sector.

\begin{figure}
    \centering
    \includegraphics[width=0.999\columnwidth]{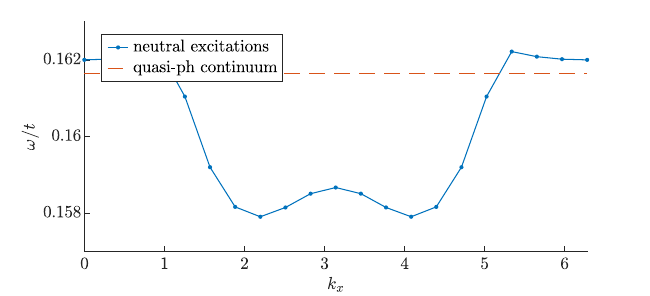}
    \caption{The dispersion relation in the charge neutral sector for the case $(\alpha,n_b)=(1/N_y,1/(2N_y))$ for $N_y=8$, as computed by the MPS excitation ansatz (blue line), compared with the edge of the quasiparticle-quasihole continuum that was determined from the domain wall excitation ansatz. Note the range on the $y$ axis in this plot, showing that the dispersion of the minimum is quite flat.}
    \label{fig:spec_ln1}
\end{figure}

Computing the quasiparticle and quasihole excitation energy as a function of momentum using the excitation ansatz, we find that both dispersion relations are essentially flat (up to very small fluctuations that seem to decrease with increasing bond dimension); this gives us a direct estimate of the quasihole and quasiparticle excitation gap. Next, we compute the dispersion relation in the charge-neutral sector using the excitation ansatz, giving us the result in Fig.~\ref{fig:spec_ln1}. Around momentum $k_x=0$ we find that the gap in the neutral sector is slightly higher than the edge of the quasiparticle-quasihole continuum. Since this is a quasiparticle-quasihole state, we expect that the excitation ansatz (which assumes a single-particle nature of the state) gives a slightly higher variational energy. For larger momenta, we find a branch with an energy that drops slightly below the continuum. This branch can be identified with the magneto-roton mode that is well-known in the continuum FQH effect~\cite{Girvin1985, Girvin1986}. In contrast to the latter case, however, we do not find a sharp minimum for this mode, which is a known consequence of the purely local interaction term we are considering~\cite{Yang2012, Repellin2014}. 

Our value of the gap at $N_y=8$ (and, therefore, \mbox{$\alpha=1/8$}), as shown in Fig.~\ref{fig:spec_ln1} is in close agreement with results from exact diagonalization on a torus geometry for the same parameters~\cite{Repellin2017}. As explained in Sec.~\ref{sec:model}, by solving the two-particle problem in the lattice model, we can relate the energy scale on the lattice to the energy scale in the continuum. Doing so, we find that our estimate of the gap converges to $\Delta \approx 0.57 \times V_0$ for increasing $N_y$, quite close to the value $0.615 \times V_0$ as reported~\cite{Regnault2003, Regnault2004, Repellin2014} directly for the continuum case.

\section{Laughlin-1/2 phase, second choice}
\label{sec:ln2}

\begin{figure}
    \centering
    \includegraphics[width=0.999\columnwidth]{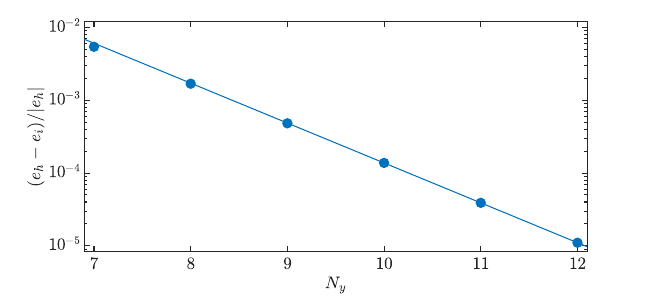}
    \includegraphics[width=0.999\columnwidth]{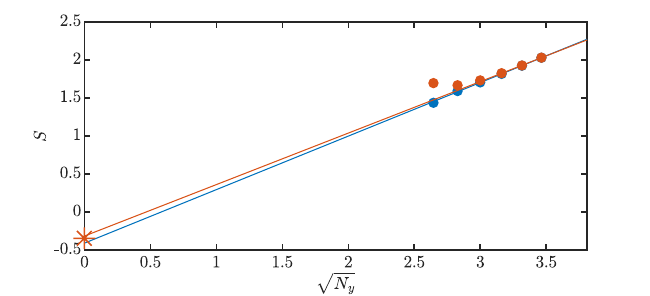}
    \caption{The scaling of the energy density difference between the two MPS ground states for the case $(\alpha,n_b)=(2/N_y,1/N_y)$ as a function of $N_y$ (top) and the scaling of the entanglement entropy of the MPS ground state in the integer and half-integer sectors (bottom) as a function of $\sqrt{N_y}$.}
    \label{fig:ln2}
\end{figure}

Keeping the magnetic filling factor fixed at $\nu=1/2$ and choosing the scaling
\begin{equation}
    \alpha=\frac{2}{N_y}, \qquad n_b=\frac{1}{N_y} \; ,
\end{equation}
we next study systems with one boson per rung, where we do not expect any breaking of the translation symmetry. Instead, in our variational MPS calculations, we consistently find two distinct uniform ground states with very small energy difference. In the top panel of Fig.~\ref{fig:ln2}, we see that the difference between the two ground-state energies in the two different sectors decreases exponentially with $N_y$. Moreover, in the bottom panel of Fig.~\ref{fig:ln2}, we see that the entanglement entropy scales with $\sqrt{N_y}$ for both ground-state sectors, showing a similar behavior as Eq.~\eqref{eq:scaling_S} and in agreement with the expected topological entanglement entropy $\gamma=\log\sqrt{2}$ for a Laughlin-1/2 phase.

\subsection{Entanglement structure}

\begin{figure}
    \centering
    \includegraphics[width=0.999\columnwidth]{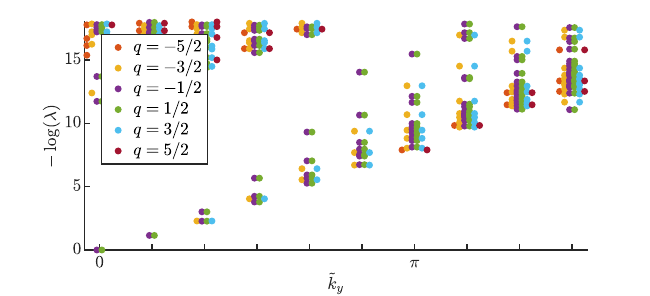}
    \includegraphics[width=0.999\columnwidth]{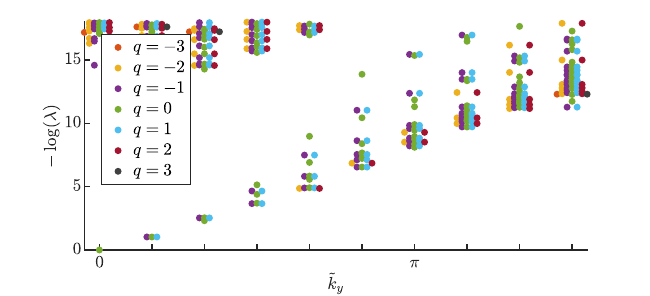}
    \caption{Entanglement spectra for the case $(\alpha,n_b)=(2/N_y,1/N_y)$ with $N_y=12$ for the two MPS ground states. The panels show the half-integer (top) and the integer sectors (bottom), where we have shifted the charge labels such that both spectra are symmetric around zero. The momentum label $\tilde{k}_y$ is defined in Eq.~\eqref{eq:ky_tilde}.}
    \label{fig:es_ln2}
\end{figure}

To understand the nature of the two different MPS ground states, we investigate their entanglement spectra. In Fig.~\ref{fig:es_ln2} we show the entanglement spectra in terms of the $\U(1)$ charge and shifted momentum labels $(q,\tilde{k}_y)$ (see Eq.~\eqref{eq:ky_tilde}) for a cylinder with $N_y=12$ sites. In the top panel, we observe non-degenerate eigenvalues in the $q=0$ sector, whereas all other sectors have identical spectra in the $q$ and $-q$ sectors. In the bottom panel, all eigenvalues are degenerate, which we implement by shifting the charges to half-integer values such that the $q\to-q$ symmetry is apparent. In this way, both spectra exhibit the structure of an effective $\OO(2)$ symmetry (as the semi-direct product of $\U(1)$ and charge conjugation), of which the integer charge sectors correspond to the faithful representations and the half-integer sectors correspond to the projective representations. This enlarged symmetry can be understood as arising from a bond-centered reflection symmetry in the MPS ground states, as noted in Sec.~\ref{sec:mps}. Because of this effective $\OO(2)$ structure, we can clearly identify the charges with the $S^z$ components of the different $\SU(2)$ multiplets of the $\SU(2)_1$ CFT (see App.~\ref{sec:app} for a more detailed comparison). As a consequence, the emergence of the multiplets is much clearer than in Fig.~\ref{fig:es_ln1_comb}.

\begin{figure}
    \centering
    \includegraphics[width=0.9999\columnwidth,page=3]{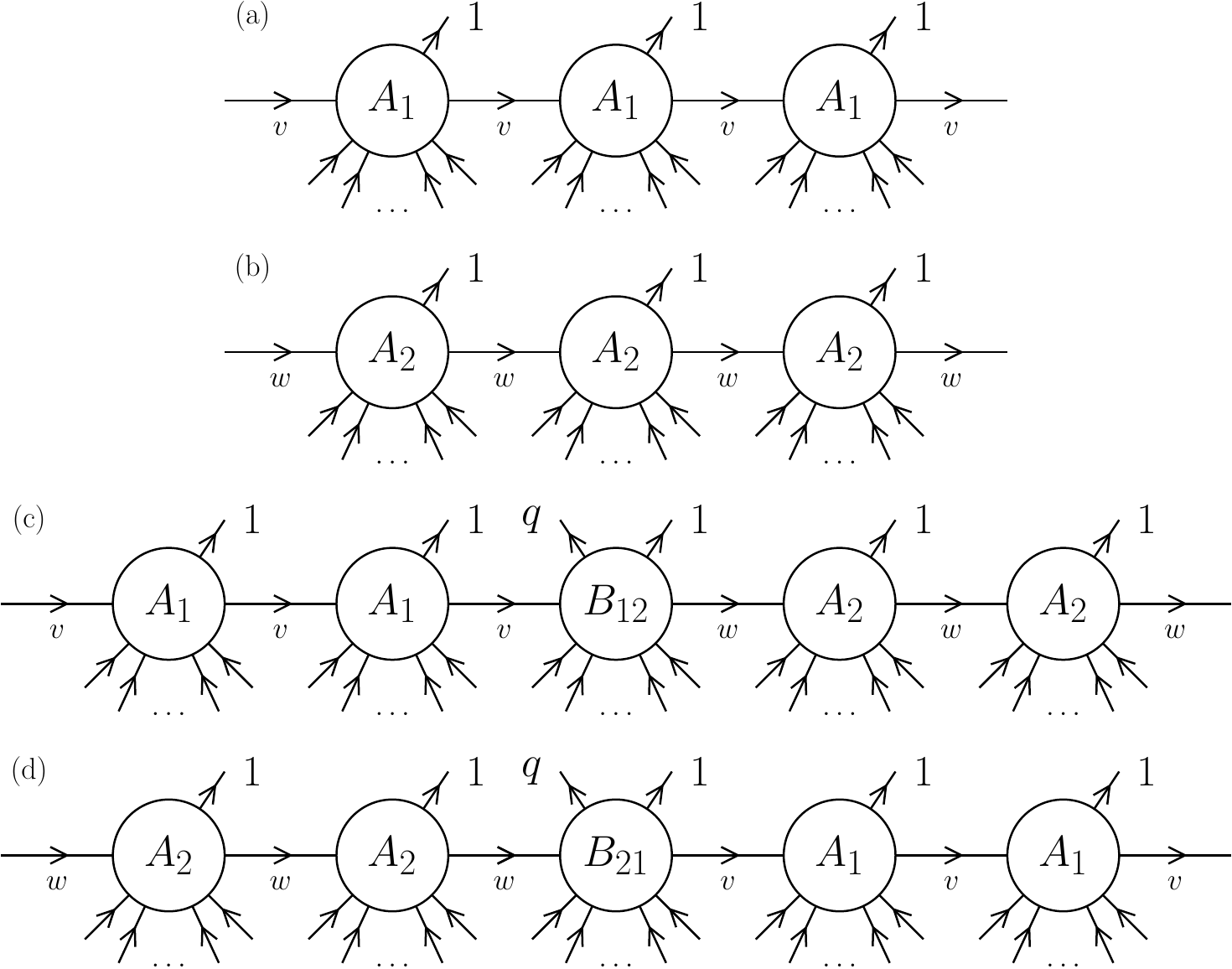}
    \caption{Diagrammatic representation of the two MPS ground states (a,b), described by super-site tensors $A_1$ and $A_2$, resp., with a uniform charge distribution of one per rung. The virtual charges denote integer and half-integer representations of the $\U(1)$ symmetry denoted by $v$ and $w$. In (c) and (d) we show the two types of domain-wall excitations, described by the domain wall tensors $B_{12}$ and $B_{21}$, which necessarily carry half-integer charges $q$ in addition to the background charge of one. Note that these domain-wall excitations are not proper excitations of the model on the cylinder: because one of the two MPS has slightly higher energies, the domain walls are confined.}
    \label{fig:mps_ln2}
\end{figure}

The occurrence of two MPS ground states with an exponentially small difference in energy follows the scenario for identifying a chiral spin liquid (CSL) phase in the context of $\SU(2)$ Heisenberg models using MPS techniques on the cylinder~\cite{Cincio2013, Bauer2014, Zaletel2016, Cincio2015}. In that context, for even-leg cylinders, one similarly finds two distinct ground states, where the virtual MPS bonds follow either half-integer or integer representations of the $\SU(2)$ symmetry. The two different MPS ground states are identified as minimally entangled states~\cite{Zhang2012, Jiang2012, Zhu2014, He2014b, Huang2014}. The structure of the entanglement spectra in Fig.~\ref{fig:es_ln2} clearly follows the same structure, with the (half-)integer representations of the effective $\OO(2)$ symmetry taking the role of the (half-)integer $\SU(2)$ representations.

The symmetry properties of these minimally entangled states are further classified in terms of symmetry-protected topological (SPT) order in one dimension~\cite{Pollmann2010, Chen2011, Schuch2011}. Indeed, as noted in Sec.~\ref{sec:mps}, the cylinder system can be interpreted as an effective 1D model, for which the anti-unitary reflection symmetry allows for an SPT $\Z_2$ invariant $\gamma=\pm1$~\cite{Zaletel2016, Cincio2015, Zaletel2017, Saadatmand2016}. As a result, any MPS ground state that preserves reflection symmetry can fall in either of the two distinct SPT classes, and two MPS in different classes cannot be adiabatically connected without breaking the symmetry. The entanglement spectrum serves as a fingerprint of the SPT classification, since the non-trivial SPT class ($\gamma=-1$) is characterized by a perfect twofold degeneracy in the MPS entanglement spectrum. Upon inspection of the entanglement spectra in Fig.~\ref{fig:es_ln2}, we can therefore associate the two different MPS ground states as belonging to the two different 1D-SPT classes. Note that the effective $\OO(2)$ structure in the entanglement spectrum leads to the similar conclusion that the two MPS cannot be adiabatically connected while conserving the $\OO(2)$ structure.

\subsection{Quantized charge pumping}
\label{sec:charge_pumping_ln2}

\begin{figure}
    \centering
    \includegraphics[width=0.999\columnwidth]{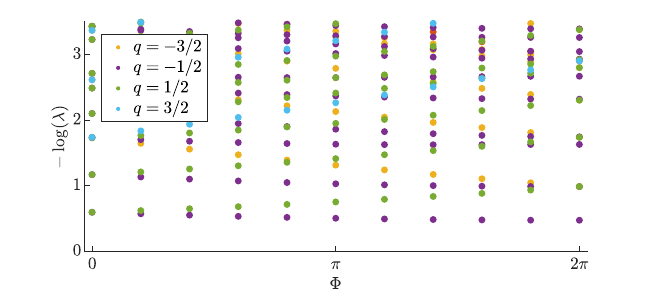}
    \caption{The entanglement spectrum as we adiabatically thread a flux $\Phi$ through the cylinder for the case $(\alpha,n_b)=(2/N_y,1/N_y)$ with $N_y=12$. We start from the MPS in the half-integer sector at $\Phi=0$. Here the charge labels are defined such that the spectrum at $\Phi=0$ is symmetric around charge zero (i.e., the spectrum in the top panel of Fig.~\ref{fig:es_ln2}), so that at $\Phi=2\pi$ the labels are shifted by one-half with respect to the spectrum in the bottom panel of Fig.~\ref{fig:es_ln2}.}
    \label{fig:flux_ln2}
\end{figure}

Similarly as in the CSL case, the two ground states can be connected adiabatically by threading a flux $\Phi$ through the cylinder, which breaks the reflection symmetry (and, therefore, the SPT classification) explicitly. In Fig.~\ref{fig:flux_ln2} we observe that the two entanglement spectra evolve continuously as a function of $\Phi$. Because of the charge conjugation symmetry in the entanglement spectrum at the starting point ($\Phi=2\pi$) and end point ($\Phi=2\pi$), $\Delta Q$ is identically zero in both spectra. Following a similar reasoning as in Sec.~\ref{sec:charge_pumping_ln1}, we deduce that this adiabatic flux insertion has transported a net charge of exactly $\Delta Q_{\rm tr}=1/2$ through the system. In contrast to the $\alpha=1/N_y$ case, here the symmetries of the states in the different topological sectors dictate that the flux insertion protocol yields an exact quantization of the Chern number $\mC=1/2$.

\subsection{Low-energy excitations}

We conclude our discussion of the Laughlin-1/2 phase with a comment on the nature of the excitations in this setting. As shown in Fig.~\ref{fig:mps_ln2}, we can again consider domain wall excitations between the two different MPS ground state approximations. Because we have assigned integer and half-integer representations to the two MPS, a domain wall again necessarily has a half-integer charge, and therefore describes quasiparticle or quasihole states. In contrast to the previous case, however, the two MPS ground states do not have the same energy. As a result, a domain wall cannot exist as a free particle, because it would have an infinite excitation energy. Instead, two domain wall states can exist as configurations on top of the lowest-energy ground state, with the higher-energy state appearing in between the two domain walls and giving rise to a linear confining potential. Therefore, we expect to see a discrete set of bound states of domain walls. As the energy difference decreases exponentially with $N_y$, this discrete spectrum is expected to become denser and eventually converge to a continuum in the 2D limit. Note that this scenario of confinement of fractionalized quasiparticles is well-known in the context of antiferromagnetic spin chains \cite{Shiba1980, Affleck1998}, and is also expected to occur as a finite-circumference effect in the spectrum of chiral spin liquids on the cylinder.

\section{Moore-Read phase}

We now repeat the same type of analysis for the Moore-Read~\cite{Moore1991} phase at $\nu=1$. This phase is known to occur in the bosonic quantum Hall problem in the continuum when three-body contact interactions are considered~\cite{Cooper1999, Wilkin2000}. The thin-cylinder analysis for the continuum FQH states (see Sec.~\ref{sec:tt}) has been extended to the $\nu=1$ case as well~\cite{Bergholtz2006b, Seidel2006}: In the limit of $L_y\to0$, the ground state subspace is now spanned by the two Tao-Thouless-like states $[2020]$ and $[0202]$, which spontaneously break translation symmetry, and the uniform state $[1111]$~\cite{Seidel2006, Liu2012}. In contrast to the Laughlin case, the formation of the crystalline states is not dictated by the Lieb-Schultz-Mattis-Oshikawa theorem, but is a consequence of the special form of the Moore-Read parent Hamiltonian in the limit $L_y\to0$~\cite{Seidel2006}. Similar to the Laughlin case, however, the CDW order parameter in the $[2020]$ and $[0202]$ states vanishes exponentially when $L_y$ grows larger. While the two crystalline states are related by translations and are necessarily degenerate, the uniform state is qualitatively different and does not have the same energy away from the small-$L_y$ limit. As shown numerically, however, the three ground states are adiabatically connected to the threefold-degenerate subspace of the $\nu=1$ Moore-Read states for large circumference~\cite{Seidel2006}. Again, these product states are the 1D precursors of the three degenerate ground states characteristic of the topological Moore-Read phase on the cylinder. 

In a number of different works, the Moore-Read phase has been identified in FCI models on the lattice~\cite{Zhu2015, Palm2021, Boesl2022, Palm2025}. Here, we consider the extended Hofstadter-Bose-Hubbard model [Eq.~\eqref{eq:ham}] with $U_2=0$ and $U_3\to\infty$, which is implemented by restricting the maximum number of bosons per site to two.

\begin{figure}
    \centering
    \includegraphics[width=0.999\columnwidth]{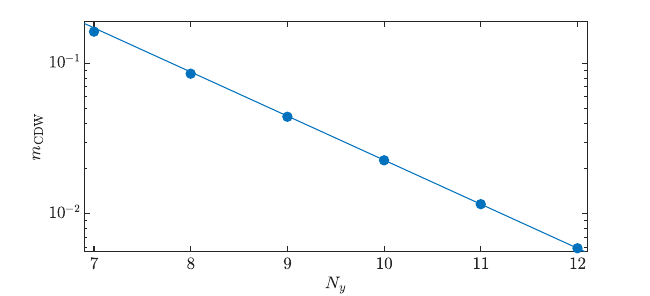}
    \caption{The CDW order parameter for the case $(\alpha,n_b)=(1/N_y,1/N_y)$ as a function of $N_y$.}
    \label{fig:cdw_mr}
\end{figure}

We first investigate the case $\alpha=1/L_y$, which amounts to setting $n_b=1/L_y$ to reach a magnetic filling factor of $\nu=1$. Since we are now working with one boson per rung, there are no obstructions for obtaining a unique gapped ground state. However, when optimizing uniform MPS, we find that the translation symmetry is broken spontaneously for all values of $L_y=7,\dots,12$. In Fig.~\ref{fig:cdw_mr}, we observe that the order parameter $m_{\mathrm{CDW}}$ [Eq.~\eqref{eq:cdw}] decreases exponentially with $L_y$, similarly to the Laughlin-1/2 case. We therefore conclude that the CDW states are the lattice equivalents of the $[2020]$ and $[0202]$ Tao-Thouless states.

For the Moore-Read phase, however, we expect a third MPS ground state, corresponding to the $[1111]$ state. Nevertheless, when variationally optimizing the MPS starting from random initial points, we cannot find this third state. We can, however, introduce a small non-zero $U_2$ favoring a uniform density. Indeed, for $U_2/t\approx1$, we can stabilize a uniform MPS ground state. Upon adiabatically decreasing the value of $U_2$ (i.e., taking small steps and always using the previous MPS as the initial state), we can find a uniform MPS ground-state approximation that is a local minimum of the variational optimization at $U_2=0$.

\begin{figure}
    \centering
    \includegraphics[width=0.999\columnwidth]{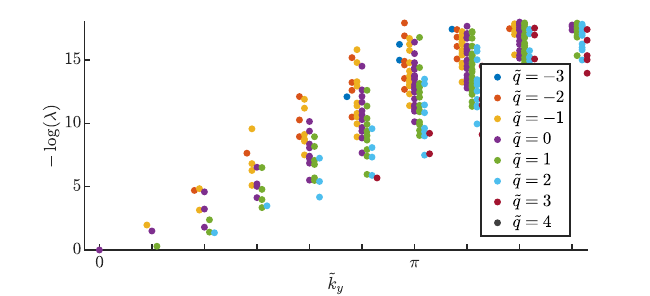}
    \includegraphics[width=0.999\columnwidth]{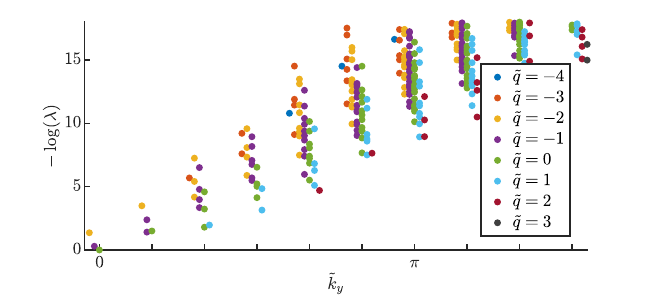}
    \includegraphics[width=0.999\columnwidth]{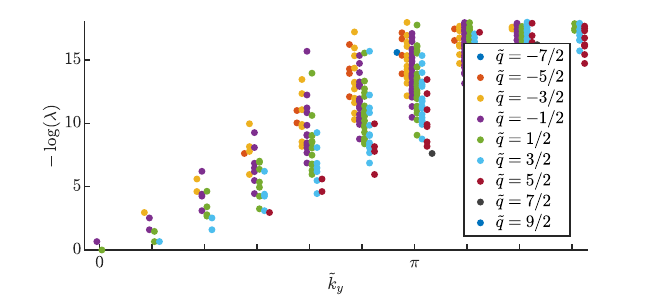}
    \caption{The entanglement spectra for the case $(\alpha,n_b)=(1/N_y,1/N_y)$ with $N_y=12$. We plot the spectra as a function of the shifted momentum $\tilde{k}_y$, with the bare charge labels $q$.}
    \label{fig:es_mr1}
\end{figure}

To better understand the nature of these three different MPS ground states, we now investigate their entanglement spectra, see Fig.~\ref{fig:es_mr1}. Just as for the CDW-ordered states in Sec.~\ref{sec:ln1}, the $[2020]$ and $[0202]$ states give rise to two entanglement spectra that are related through charge conjugation (up to shifts in the momentum $k_y$). Upon shifting the momentum through Eq.~\eqref{eq:ky_tilde}, the $\tilde{k}_y$ resolved entanglement spectra in the top and middle panels of Fig.~\ref{fig:es_mr1} show the counting of the vacuum and fermion sectors of the $\SU(2)_2$ chiral CFT (see App.~\ref{sec:app} for more details). Similar to the entanglement spectra in Sec.~\ref{sec:ln1}, the splittings within the multiplets are necessarily very large, because they are related through charge conjugation.

The $[1111]$ state is a uniform state, resulting in a spectrum with an effective charge conjugation symmetry. The spectrum is centered around $q=0$, which, in the 1D-SPT language of Sec.~\ref{sec:ln2}, amounts to a trivial phase. When we plot the spectrum as a function of the shifted momentum $\tilde{k}_y$, however, we find the counting to match the half-integer Majorana sector of the chiral $\SU(2)_2$ CFT. In Fig.~\ref{fig:es_mr1} we therefore shifted the charge labels to half-integer values $\tilde{q}$ to obtain the right multiplet structure. At first sight, for the spectrum to match the half-integer CFT sector, it would be a lot more natural if the $[1111]$ state lived in the non-trivial 1D-SPT sector with a purely twofold degenerate spectrum. We note, however, that the Tao-Thouless states are expected to be adiabatically connected to a product state, and can therefore necessarily only transform trivially under the anti-unitary reflection symmetry.

\begin{figure}
    \centering
    \includegraphics[width=0.999\columnwidth]{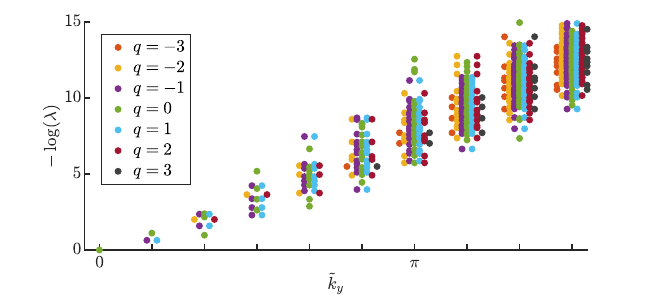}
    \includegraphics[width=0.999\columnwidth]{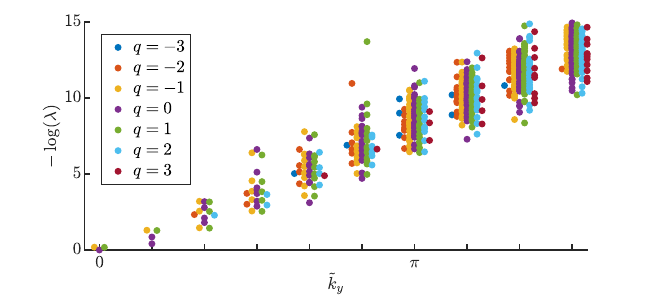}
    \caption{The entanglement spectra for the case $(\alpha,n_b)=(2/N_y,2/N_y)$ with $N_y=12$. We plot the spectra as a function of the shifted momentum $\tilde{k}_y$ and the bare charge labels $q$.}
    \label{fig:es_mr2}
\end{figure}

Finally, we consider the case $\alpha=2/N_y$, requiring a boson density of $n_b=2/N_y$ to reach the magnetic filling factor $\nu=1$, i.e., two bosons per rung, where again we do not have any obstructions to have an MPS ground state with a single unit cell. When we optimize an MPS ground state approximation for this setting at $N_y=12$, we consistently find an MPS with a charge-conjugate symmetric entanglement spectrum, shown in the top panel of Fig.~\ref{fig:es_mr2}. The counting corresponds to the one of the vacuum sector of the $\SU(2)_2$ CFT.

A second MPS can be reached using the flux threading procedure. The entanglement spectrum for this state exhibits a counting consistent with the fermionic sector of the $\SU(2)_2$ CFT (bottom panel of Fig.~\ref{fig:es_mr2}). Both spectra are symmetric around the $q=0$ sector with integer charge labels, so they are both trivial states in the 1D-SPT classification from Sec.~\ref{sec:ln2}. Following the reasoning from Sec.~\ref{sec:charge_pumping_ln2}, the flux-threading protocol gives us a transported charge of exactly $Q_{\mathrm{tr}}=1$, in agreement with a many-body Chern number $\mC=1$ for the Moore-Read state.

Also for the case $\alpha=2/N_y$ we expect to find a third state representing the Majorana sector. Yet, in our numerical MPS optimizations, we have not been able to find this state as a local variational optimum (also upon addition of stabilizing terms in the Hamiltonian). From all our previous observations, it is clear that we expect this state to have an entanglement spectrum symmetric around $q=0$ with half-integer labels, belonging to the non-trivial 1D-SPT class. Note that such an entanglement spectrum (along with the two other sectors) was found in infinite MPS simulations on the cylinder for an FCI model on the honeycomb lattice with next-nearest-neighbor hoppings that further flatten the band dispersion~\cite{Zhu2015}.

\section{Conclusions and outlook}

In this work, we have studied the combination of finite-circumference and lattice effects on the topological signatures of FCI states in the Hofstadter-Bose-Hubbard model on infinite cylinders. To this end, we have optimized MPS ground-state approximations for an increasing number of sites along the cylinder $N_y$, while also scaling the flux per plaquette as $\alpha\propto1/N_y$. We have focused on the symmetry properties of the entanglement spectrum as the prime signature for diagnosing the topological order. 

If we choose to scale the flux as $\alpha=1/N_y$, we find lattice equivalents of Tao-Thouless states, both for the Laughlin-1/2 ($\nu=1/2$) and the Moore-Read phase ($\nu=1$). In particular, we find that the ground state develops CDW order, which decays exponentially with $N_y$. In the Laughlin-1/2 case, the CDW ordering in the $[1010]$ and $[0101]$ states is a direct consequence of the Lieb-Schultz-Mattis-Oshikawa theorem, whereas in the Moore-Read case, the two CDW-ordered states $[2020]$ and $[0202]$ appear spontaneously, in addition to the uniform state [1111]. As a result of the spontaneous breaking of spatial symmetries in these ground states on the cylinder, in the entanglement spectra we have observed large deviations from the $\SU(2)$ multiplet structure that is expected from the emergent CFT description of the edge.

If we choose to scale the flux as $\alpha=1/(2N_y)$, we find a very different situation without a direct analogue in the continuum. For the Laughlin-1/2 case, we find two ground states with an energy density difference that decays exponentially with $N_y$. This situation is entirely analogous to the classification of the topological ground-state degeneracy of chiral spin liquids on infinite cylinders in terms of minimally entangled states and 1D-SPT invariants (where here the effective $\OO(2)$ structure in the entanglement spectrum takes the role of the $\SU(2)$ symmetry in the spin-liquid case). In this case, because the spatial symmetries remain unbroken in the ground states, we could observe the approximate formation of the $\SU(2)$ multiplet structure in the entanglement spectrum. The Moore-Read case again follows a similar analysis, but using variational MPS simulations, we have only been able to stabilize two of the three topological sectors. A possible avenue for finding a good initial point for the variational MPS optimization of the Majorana sector would be the conversion of projected fermionic Gaussian states into MPS~\cite{Wu2020, Liu2025}.

Looking forward, it would be promising to understand the precise nature of the finite-circumference corrections to the entanglement spectra that we have observed in our work. For that purpose, it would be worthwhile to extend the formalism that was developed for fitting finite-size entanglement spectra of CSL states~\cite{Arildsen2022} and FQH states \cite{Arildsen2025} to the context of bosonic lattice systems, where we have shown that the $\SU(2)$ multiplet structure is further broken down to an effective $\OO(2)$ structure.

\section{Acknowledgments}
We thank Botao Wang and Nathan Goldman for collaborations on a related work and Julian Boesl and C\'ecile Repellin for interesting discussions. LV would like to thank Atsushi Ueda and Mark Arildsen for helping out with the CFT counting, and Jheng-Wei Li and Hong-Hao Tu for inspiring discussions. We acknowledge financial support by the ERC Consolidator Grant LATIS, and the EOS Project CHEQS.

\bibliography{bibliography.bib}

\appendix
\section{Details on CFT counting}
\label{sec:app}

\begin{figure*}
    \centering
    \includegraphics[scale=0.8]{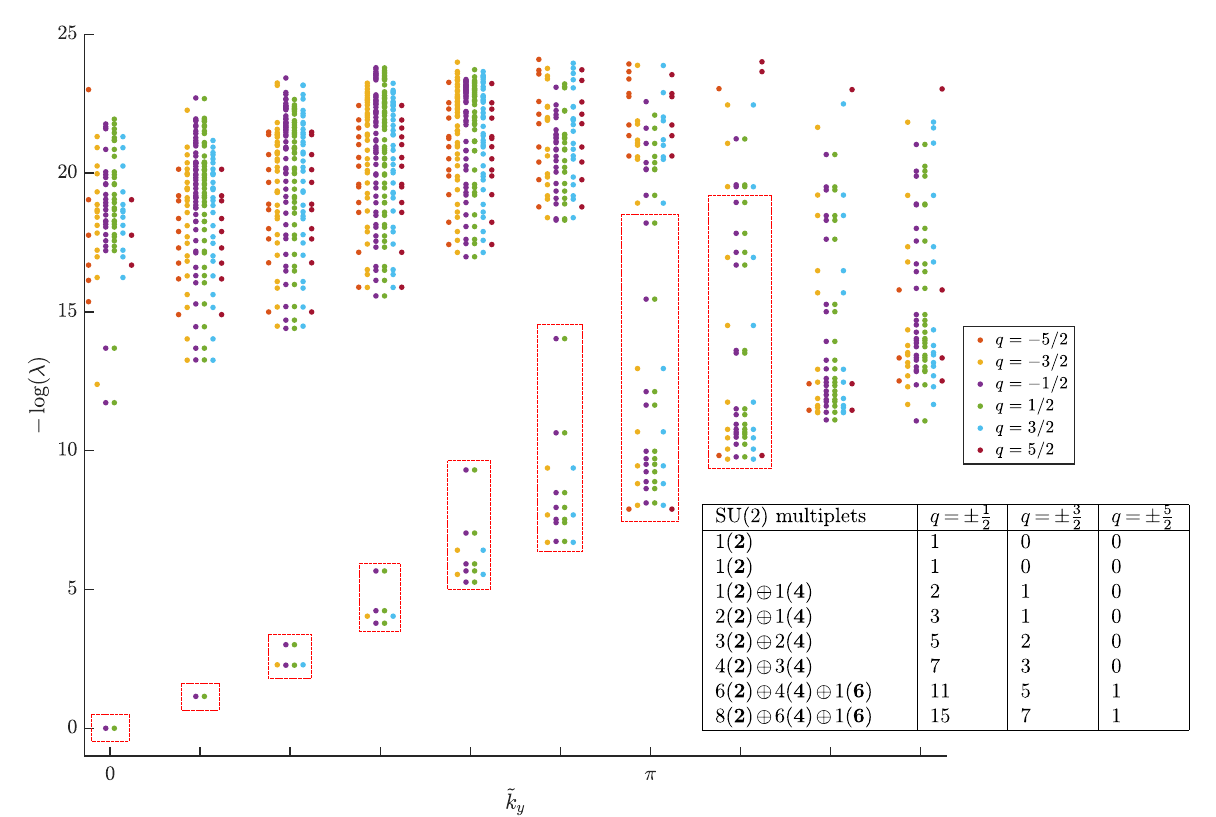}
    \includegraphics[scale=0.8]{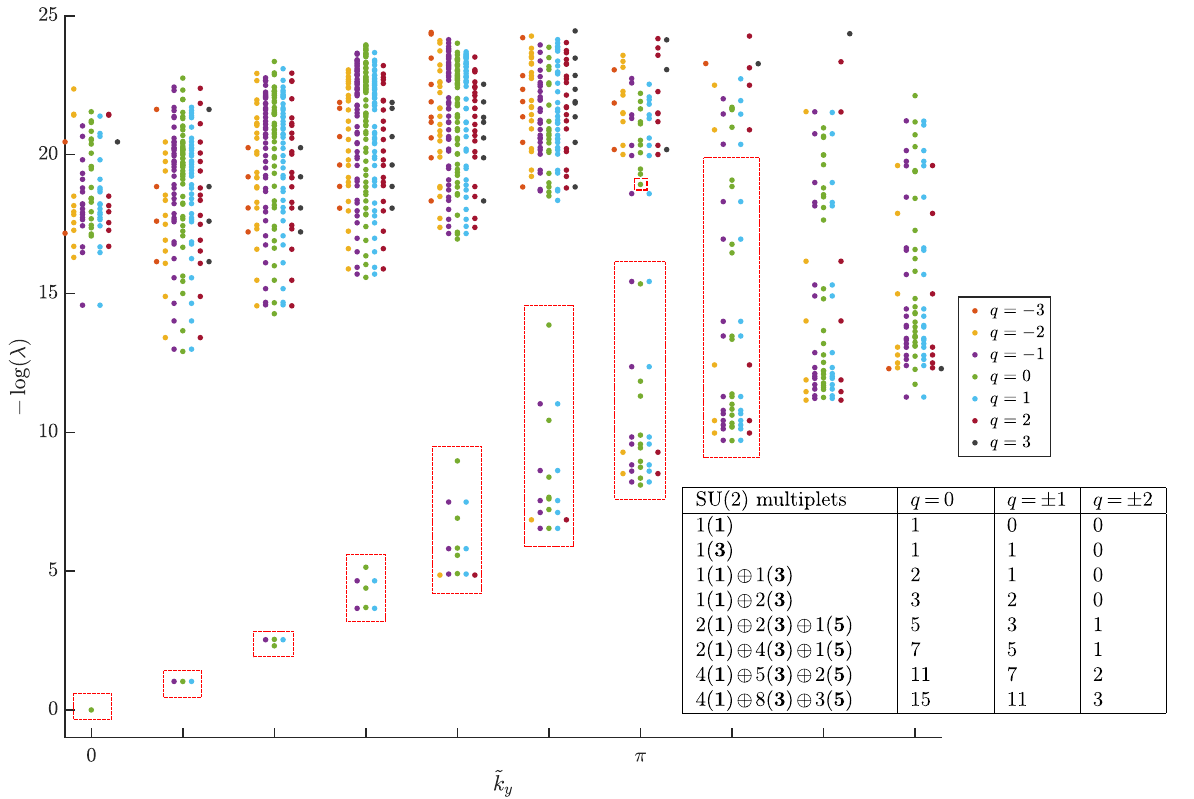}
    \caption{More detailed plot of the entanglement spectrum from Fig.~\ref{fig:es_ln2}, this time compared explicitly with the expected $\SU(2)_1$ CFT counting. The tables show the first levels that are expected from the chiral CFT, as found in Ref.~\onlinecite{Kass1990}, along with the expected total 
    degeneracies in the different $q$ sectors. The red boxes show which Schmidt values we have to take in order to match the CFT counting in the tables (for the higher levels there is no longer a clear gap to the non-universal part of the spectrum, so the boxes are a bit voluntaristic in this regime).}
    \label{fig:es_ln2_with_boxes}
\end{figure*}

In this appendix we give the tables for the level counting of the chiral $\SU(2)_k$ Wess-Zumino-Witten CFT for $k=1$ (Tab.~\ref{tab:su2_1}) and $k=2$ (Tab.~\ref{tab:su2_2}). For more details on the derivation we refer to specialized works; here we reproduce the tables from Ref.~\onlinecite{Kass1990}. The different levels can be decomposed into SU(2) multiplets, which we denote by their dimension, i.e. $\one$ for the trivial representation, $\two$ for the half-integer representation, etc. Furthermore, in Fig.~\ref{fig:es_ln2_with_boxes} we show a close-up of the entanglement spectra from Fig.~\ref{fig:es_ln2}, where we show the agreement with the $\SU(2)_1$ CFT counting in detail.

\begin{table*}
\begin{tabular}{|l|l|l|}
        \hline
        & $j=0$ & $j=1$ \\
        \hline
        $k=0$  & $ 1  \one$                                                      & $ 1  \two$              \\
        $k=1$  & $ 1  \three$                                                    & $ 1  \two$              \\
        $k=2$  & $ 1  \one \oplusL 1  \three$                                    & $ 1  \two \oplusL 1  \four $      \\
        $k=3$  & $ 1  \one \oplusL 2  \three$                                    & $ 2  \two \oplusL 1  \four $     \\
        $k=4$  & $ 2  \one \oplusL 2  \three \oplusL 1  \five $                  & $ 3  \two \oplusL 2  \four $    \\
        $k=5$  & $ 2  \one \oplusL 4  \three \oplusL 1  \five $                  & $ 4  \two \oplusL 3  \four $ \\
        $k=6$  & $ 4  \one \oplusL 5  \three \oplusL 2  \five $                  & $ 6  \two \oplusL 4  \four \oplusL 1 \five $\\
        $k=7$  & $ 4  \one \oplusL 8  \three \oplusL 3  \five $                  & $ 8  \two \oplusL 6  \four \oplusL 1 \five $\\
        $k=8$  & $ 7  \one \oplusL 10 \three \oplusL 5  \five $                  & $ 11 \two \oplusL 9  \four \oplusL 2 \five $\\
        $k=9$  & $ 8  \one \oplusL 15 \three \oplusL 6  \five \oplusL 1 \seven $ & $ 15 \two \oplusL 12 \four \oplusL 3 \five $\\
        $k=10$ & $ 12 \one \oplusL 19 \three \oplusL 10 \five \oplusL 1 \seven $ & $ 20 \two \oplusL 17 \four \oplusL 5 \five $\\
        $k=11$ & $ 14 \one \oplusL 27 \three \oplusL 13 \five \oplusL 2 \seven $ & $ 26 \two \oplusL 23 \four \oplusL 7 \five $\\
        \hline
    \end{tabular}
    \caption{Table for the first eleven levels of the $\SU(2)_1$ CFT countings in the two primary sectors (also called integer and half-integer sectors in the main text).}
    \label{tab:su2_1}
\end{table*}
\begin{table*}
    \begin{tabular}{|l|l|l|l|}
        \hline
        & $j=0$ & $j=1/2$ & $j=1$  \\
        \hline
        $k=0$  & 1$\one$                                                            & $1\two$                                           & $1 \three$                 \\
        $k=1$  & 1$\three$                                                          & $1\two\oplusL1\four$                              & $1\one\oplusL1\three$             \\
        $k=2$  & $1\one\oplusL1\three\oplusL1\five$                                 & $2\two\oplusL2\four$                              & $1\oplusL2\three\oplusL\five$          \\
        $k=3$  & $1\one\oplusL3\three\oplusL\five$                                  & $4\two\oplusL3\four\oplusL1\six$                  & $2\one\oplusL3\three\oplusL2\five$      \\
        $k=4$  & $3\one\oplusL4\three\oplusL3\five$                                 & $6\two\oplusL6\four\oplusL2\six$                  & $3\one\oplusL6\three\oplusL3\five\oplusL1\seven$  \\
        $k=5$  & $3\one\oplusL8\three\oplusL4\five\oplusL\seven$                    & $10\two\oplusL10\four\oplusL4\six$                & $5\one\oplusL9\three\oplusL6\five\oplusL1\seven$             \\
        $k=6$  & $7\one\oplusL11\three\oplusL8\five\oplusL2\seven$                  & $16\two\oplusL16\four\oplusL7\six\oplusL1\eight$  & $7\one\oplusL15\three\oplusL10\five\oplusL3\seven$ \\
        $k=7$  & $8\one\oplusL19\three\oplusL12\five\oplusL4\seven$                 & $24\two\oplusL26\four\oplusL12\six\oplusL2\eight$ & $12\one\oplusL22\three\oplusL16\five\oplusL5\seven$ \\
        $k=8$  & $15\one\oplusL27\three\oplusL21\five\oplusL6\seven\oplusL1\seven$  & $36\two\oplusL40\four\oplusL20\six\oplusL4\eight$ & $16\one\oplusL35\three\oplusL25\five\oplusL9\seven\oplusL1\nine$ \\
        $k=9$  & $19\one\oplusL43\three\oplusL30\five\oplusL12\seven\oplusL1\seven$ & $54\two\oplusL60\four\oplusL32\six\oplusL8\eight$ & $25\one\oplusL50\three\oplusL39\five\oplusL14\seven\oplusL2\nine$ \\
        \hline
    \end{tabular}
    \caption{Table for the first nine levels of the $\SU(2)_2$ CFT countings in the three primary sectors (also called vacuum, Majorana and fermion sectors in the main text).}
    \label{tab:su2_2}
\end{table*}

\end{document}